\documentclass[a4paper,12pt,parskip=half]{scrarticle}

\usepackage{amsmath}
\usepackage{amssymb}
\usepackage[affil-it]{authblk}
\usepackage{booktabs}
\usepackage{geometry}
\usepackage{graphicx}
\usepackage{microtype}
\usepackage{pifont}
\usepackage{setspace}
\usepackage{xcolor}
\usepackage{helvet}
\usepackage{siunitx}
\usepackage[T1]{fontenc}
\usepackage{hyperref}

\newcommand{\centeredgraphics}[2][]{\vcenter{\hbox{\includegraphics[#1]{#2}}}}

\graphicspath{{figures/}{build/figures/}}

\definecolor{desycyan}{rgb}{0,0.623,0.874}

\hypersetup{
  colorlinks=true,
  allcolors=desycyan,
  linkcolor=desycyan,
  citecolor=desycyan,
  filecolor=desycyan,
  menucolor=desycyan,
  runcolor=desycyan,
  urlcolor=desycyan
}

\begin{document}

\title{Scaling up to Multivariate Rational Function Reconstruction}
\author{\normalsize Andreas Maier}
\affil{\small Deutsches Elektronen-Synchrotron DESY, Platanenallee 6,
  15738 Zeuthen, Germany}
\date{}

\maketitle

\begin{abstract}
  \noindent I present an algorithm for the reconstruction of multivariate rational
  functions from black-box probes. The arguably most important
  application in high-energy physics is the calculation of multi-loop
  and multi-leg amplitudes, where rational functions appear as
  coefficients in the integration-by-parts reduction to basis
  integrals. I show that for a dense coefficient the algorithm is nearly
  optimal, in the sense that the number of required probes is close to
  the number of unknowns.
\end{abstract}

\section{Introduction}
\label{sec:intro}

The reduction of a large number of scalar Feynman integrals to a
smaller set of basis (or master) integrals is an almost universal step in
precision calculations in quantum field theories. In many cases, it is
also among the most challenging parts of the computation, and has
therefore seen lots of attention and development over the years.

The current standard approach is to derive integration-by-parts
identities~\cite{Chetyrkin:1981qh,Tkachov:1981wb} for a set of seed integrals with
fixed powers of propagators and irreducible scalar products and solve
the resulting system of linear relations via Gauss
elimination~\cite{Laporta:2001dd}. The result is a subset of the seed
integrals, each expressed in terms of a linear combination of basis
integrals. The coefficients are rational functions of kinematic
invariants and the space-time dimension.

It is often advantageous to insert numerical values for the dimension
and the invariants and solve the system over a finite
field~\cite{Kauers:2008zz}. This strategy was initially used to
quickly eliminate redundant relations~\cite{Kant:2013vta}. However,
solving the system with sufficiently many different probes, i.e.\
different numeric values for the variables, it is possible to
reconstruct the full
result~\cite{vonManteuffel:2014ixa,Peraro:2016wsq}. In this way, one
avoids large intermediate expressions and can restrict the
reconstruction to those coefficients that are actually needed for the
final result, for example a scattering amplitude. Further advantages
include ease of parallelisation, lower memory usage, and the
possibility to optimise the system further after a computationally
cheap pilot run.

This general strategy has been further developed along several
directions. Right from the start, one can attempt to find combinations
of relations that lead to better behaved systems or even an explicit
recursive solution to the reduction
problem~\cite{Gorishnii:1989gt,Steinhauser:2000ry,Gluza:2010ws,Schabinger:2011dz,Lee:2012cn,Lee:2013mka,Ita:2015tya,Larsen:2015ped,Ruijl:2017cxj,Pikelner:2017tgv,Wu:2023upw,Guan:2024byi}. Since
in a finite-field reduction the very same system of linear equations
has to be solved many times, it can be worthwhile to record the
solution steps once, optimise the recorded sequence, and replay it to
rapidly obtain further probes~\cite{Magerya:2022hvj}. Other efforts
target the reconstruction itself. The complexity of the rational
function coefficients depends on the chosen basis integrals, and a
judicious choice results in a factorisation between the dimension and
the kinematic
invariants~\cite{Smirnov:2020quc,Usovitsch:2020jrk}. Identifying
common factors in the coefficients can further reduce the number of
probes required for the
reconstruction~\cite{Abreu:2018zmy,DeLaurentis:2022otd,Chawdhry:2023yyx,Liu:2023cgs}. Finally,
the number of required probes depends on the chosen reconstruction
method, and various algorithms with different strengths have been
explored~\cite{Klappert:2019emp,Peraro:2019svx,Klappert:2020aqs,Belitsky:2023qho}.

In the following, I present an algorithm for ``scaling up'' from
rational function reconstruction in a single variable to the
multivariate case. In section~\ref{sec:uni_rec}, I review
reconstruction in a single variable using Thiele
interpolation~\cite{thiele}. I then discuss a way to generalise the
method to the multivariate case in section~\ref{sec:multi_rec}. The
intended application is the reconstruction of coefficients in the
reduction to basis integrals. In section~\ref{sec:application}, I
apply the algorithm to complex reduction coefficients in a massive
four-loop propagator example and in the two-loop amplitude for diphoton plus
jet production~\cite{Agarwal:2021vdh}. I find that the number of
required probes is close to optimal when the fraction of vanishing
polynomial coefficients in the numerator and the denominator of the
rational function is small.

\section{Univariate Rational Function Reconstruction}
\label{sec:uni_rec}

Excellent introductions into the reconstruction of polynomials and
rational functions are given in
\cite{Peraro:2016wsq,Klappert:2019emp}. Let us briefly review the case
of a univariate rational function.

We are given a rational function $f$ in a single variable for which we
want to find an explicit form
\begin{equation}
  \label{rat_fun}
  f(x) = \frac{p_n(x)}{p_d(x)},
\end{equation}
where $p_n, p_d$ are unknown polynomials with no common roots. $f$ is
a ``black box'', meaning that our only piece of information is an
algorithm for computing $f(t)$ for any $t$ in the domain of
$f$. One strategy is to construct rational interpolations $f_N$ for
$N$ probes $(t_i, f(t_i))$ with $i=1,\dots,N$. If $N$ is large enough,
we then find $f_N = f$ with high probability.

We start with a single probe $(t_1, f(t_1))$ and a constant interpolation
\begin{equation}
  \label{f_1}
  f_1(x) = a_1.
\end{equation}
Requiring $f_1(t_1) = f(t_1)$ we immediately find $a_1 = f(t_1)$. We
then add a second probe $(t_2, f(t_2))$. If $f_1(t_2) = f(t_2)$ we
note that we found agreement and continue with the next
probe. Otherwise, we introduce the interpolation
\begin{equation}
  \label{f_2}
  f_2(x) = a_1 + \frac{x-t_1}{a_2}
\end{equation}
with $a_1=f(t_1)$ as before and
$a_2 = \frac{t_2 - t_1}{f(t_2) - a_1}$. In general, after $N$
independent probes the interpolation has the form~\cite{thiele}
\begin{equation}
  \label{f_N}
  f_N(x) = a_1 + \frac{x - t_1} {a_2 + \frac{x - t_2}{ a_3 + \frac{x - t_3}{\dots + \frac{x - t_{N-1}}{a_N}}}}.
\end{equation}
For adding the next probe $(t_{N+1}, f(t_{N+1}))$ we compute
\begin{equation}
  \label{eq:a_frac_recursion}
  c_1 = f(t_{N+1}), \qquad c_{i+1} = \frac{t_{N+1} - t_i}{c_i - a_i},
\end{equation}
for all $i \le N$ and construct $f_{N+1}$ with $a_{N+1} = c_{N+1}$ and $a_1,\dots,a_N$
taken from $f_N$. If the denominator in
equation~\eqref{eq:a_frac_recursion} vanishes, the probe adds no new
information. We then check if our present interpolation already agrees
for this point, i.e.\ $f_{N}(t_{N+1}) = f(t_{N+1})$, and terminate the
reconstruction as soon as some chosen number of probes are predicted
correctly. Otherwise we continue with the next probe.

In many cases, the computational cost of the reconstruction is
dominated by either the evaluation of the probes or by the divisions
in the calculation of the auxiliary constants $c_j$ in
equation~\eqref{eq:a_frac_recursion}. In the latter case, the
alternative division-free recursion
\begin{align}
  \label{eq:a_recursion_n}
  n_1 ={}& f(t_{N+1}), &  n_{i+1} ={}& (t_{N+1} - t_i) d_i, \\
  \label{eq:a_recursion_d}
  d_1 ={}& 1, &  d_{i+1} ={}& n_i - a_i d_i,
\end{align}
leading to $a_{N+1} = \frac{n_{N+1}}{d_{N+1}}$ can be much more efficient.

Note that the reconstruction is optimal if two conditions are
fulfilled. First, the degrees of the numerator and the denominator in
$f$ should be equal or the degree of the numerator should be larger
by one. Second, the numerator and denominator polynomials should be
perfectly dense, with all coefficients non-zero. In this case the
number of required probes is equal to the number of unknown
coefficients plus the chosen number of probes used for
confirmation. Empirically, the rational functions encountered in
integration-by-parts reduction without any kinematic invariants are
close to ideal with a typical overhead of about 10\% in the number of
required probes.

\section{Scaling up to Multiple Variables}
\label{sec:multi_rec}

In general, the rational function to be reconstructed has the form
\begin{equation}
  \label{f_multivar}
  f(x_1,\dots,x_n) = \frac{
    \sum C_{p_1,\dots,p_n} x_1^{p_1}\cdots x_n^{p_n}
  }{
    \sum D_{q_1,\dots,q_n} x_1^{q_1}\cdots x_n^{q_n}
  },
\end{equation}
where the sums run over all powers
$0\leq p_1 \leq P_1,\dots,0\leq p_n \leq P_n$ in the numerator and
$0\leq q_1 \leq Q_1,\dots,0\leq q_n \leq Q_n$ in the denominator. The
degrees $P_1,\dots,P_n$ and $Q_1,\dots,Q_n$ are a priori unknown. To
reconstruct a pair $(P_i, Q_i)$ of degrees we can set all other
variables to some fixed value, i.e.\ $x_j = t_j$ for all $j \ne i$,
and perform a univariate rational function reconstruction in the
remaining free variable $x_i$~\cite{Peraro:2016wsq,Klappert:2019emp,CuytL11}.

Once we know the degrees, we can in principle determine the unknown
coefficients $C, D$ by simply solving a linear system of
equations. Knowing the value of $f(t_1, \dots, t_n)$, we obtain the
linear equation
\begin{equation}
  \sum C_{p_1,\dots,p_n} t_1^{p_1}\cdots t_n^{p_n} = f(t_1, \dots, t_n) \sum D_{q_1,\dots,q_n} t_1^{q_1}\cdots t_n^{q_n}
\end{equation}
directly from the definition in equation~\eqref{f_multivar}. For $N+1$
coefficients $C_{p_1,\dots,p_n}, D_{q_1,\dots,q_n}$, we require $N$
probes to express them in terms of a single coefficient which we can
set to an arbitrary non-zero value to fix the overall
normalisation. If the numerator and denominator polynomials are dense,
the reconstruction is optimal in the sense of needing the lowest
possible number of probes. However, assuming constant-time arithmetic
for the arguments and coefficients, the time complexity for solving
the dense linear system is $\mathcal{O}(N^3)$ with a space complexity
of $\mathcal{O}(N^2)$. In practice it is usually better to use a
method that requires more probes but has better scaling behaviour.

The univariate reconstruction based on Thiele interpolation we
discussed in section~\ref{sec:uni_rec} only requires $\mathcal{O}(N)$
space to store the arguments $t_1,\dots,t_N$ and the coefficients
$a_1,\dots,a_N$. To determine these coefficients, one needs to
calculate $N (N+1)/2 = \mathcal{O}(N^2)$ auxiliary coefficients
(cf.~equation~\eqref{eq:a_frac_recursion}), each of which can be
computed in constant time. Can we generalise that method to the
multivariate case while preserving the superior scaling behaviour?

The main idea is to set all of the variables $x_1,\dots,x_n$ to a
single variable $x$, scaled to distinct powers so that we can recover
the full dependence on $x_1,\dots,x_n$ after the
reconstruction. Concretely, we consider the auxiliary function
$g(x) = f(x^{\alpha_1},\dots,x^{\alpha_n})$ with
\begin{equation}
  \label{eq:scaling_rec}
  \alpha_1 = 1, \qquad \alpha_{i+1} = [1 + \max(P_i, Q_i)] \alpha_i.
\end{equation}
We then use univariate reconstruction in $x$ to find an explicit form
for $g(x)$. For each term of the form $C_i x^i$ we can formally
interpret the power $i$ as a number in a mixed radix numeral system,
where the individual digits correspond to the powers $p_1,\dots,p_n$
of $x_1,\dots, x_n$.

Let us consider a simple example for a black-box function
$f(x_1, x_2)$.  Setting $x_2$ to a fixed value $t_2$ and using
univariate reconstruction in $x_1$ we find
\begin{equation}
  f(x_1, t_2) = \frac{C_0(t_2) + C_1(t_2) x_1}{D_0(t_2) + D_2(t_2) x_1^2}.
\end{equation}
We ignore the coefficients depending on $t_2$; our only goal was to
learn that the largest power of $x_1$ is 2. This tells us to set
$\alpha_2 = 2 + 1$, so we introduce the auxiliary function
$g(x) = f(x, x^3)$. From univariate reconstruction we obtain
\begin{equation}
  g(x) = \frac{1+x+x^4}{1+x^2+x^3}.
\end{equation}
The last step is to read off the corresponding powers of the original
variables $x_1, x_2$. For the exponents in $g$ we have mixed radix notation
$1 = 0_?1_3, 2 = 0_?2_3, 3= 1_?0_3, 4 = 1_?1_3$, where the subscript
indicates the numeral base of the corresponding position and the base
of the leading digit is irrelevant. This tells us that the original
function is
\begin{equation}
  f(x_1, x_2) = \frac{1 + x_1 + x_1 x_2}{1 + x_1^2 + x_2}.
\end{equation}

The method described so far can suffer from accidental cancellations
between numerator and denominator. For example, for
$f(x_1, x_2) = \frac{x_2}{x_1}$ we would obtain the auxillary function
$g(x) = x$, which would lead us to believe the original function was
$f(x_1, x_2) = x_1$. To prevent this, we additionally shift the
rescaled argument by a randomly chosen number.\footnote{%
  This random shift is also used with a slightly different purpose in
  the reconstruction algorithm by Cuyt and
  Lee~\cite{Peraro:2016wsq,Klappert:2019emp,CuytL11}. There, the goal
  is to ensure a unique structure and uniform coefficient
  normalisation of the reconstructed function.} Spurious cancellations
have to involve two or more different variables. We therefore expect
to avoid them by having at least one shifted variable in each possible
combination, i.e.\ we shift each variable except one.

Let us summarise the algorithm. Given a rational black-box function $f$ in $n$
variables $x_1,\dots,x_n$
\begin{enumerate}
\item For each variable $x_i$ with $i < n$, find the largest powers
  $P_i$ and $Q_i$ in the numerator and denominator. To do this, set
  all other variables to randomly chosen values, $x_j = t_j$ for all
  $j\ne i$, and use univariate reconstruction in $x_i$.
\item Compute the scaling powers $\alpha_1,\dots,\alpha_n$ using
  equation~\eqref{eq:scaling_rec} and choose random shifts
  $s_1,\dots,s_n$. One of the shifts can be set to zero, e.g.\
  $s_1=0$.
\item Use univariate reconstruction to find $g(x)$ from probes
  $(t_i, f(t_i^{\alpha_1} + s_1,\dots,t_i^{\alpha_n} + s_n))$.
\item For each term $C_i x^i$ in $g(x)$, recover the powers $p_1,\dots,p_n$
  of the original variables $x_1,\dots,x_n$ from the mixed-radix digits of $i$. Then, replace $x^i \to (x_1 - s_1)^{p_1} \cdots (x_n - s_n)^{p_n}$.
\end{enumerate}

For the main application we have in mind, namely the reconstruction of
coefficients in the reduction to basis integrals, one aims to recover
many rational functions from probes with the same arguments. One way
to use the same arguments
$t_i^{\alpha_1} + s_1,\dots,t_i^{\alpha_n} + s_n$ for different
functions $f,h,\dots$ is to choose the exponents
$\alpha_1,\dots,\alpha_n$ according to the maximum powers of the
respective variables in \emph{any} of the numerators and denominators
of $f,h,\dots$. Often, the highest powers of all variables will be
determined by a single function, such that the maximum number of
required probes remains unaffected. The price to pay is that for the
simple functions many vanishing coefficients will be
reconstructed, increasing the computing time required for the
reconstruction itself. Alternatively, different sets of probe
arguments can be used for functions of widely disparate
complexity.

One of the main advantages of numerical reduction is ease of
parallelisation. This requires that subsequent probes can be chosen
without having to wait for the outcome of feeding earlier probes into
the reconstruction algorithm. In this respect, algorithms for the
reconstruction of dense rational functions tend to perform better than
methods aimed at sparse rational functions with many vanishing
coefficients, as for the latter the probe selection typically has
to be adjusted dynamically. The presented algorithm mostly decouples
the seed choice from the reconstruction progress. For $n$ variables,
the selection strategy has to be updated $n$ times --- after
determining the powers of each of the first $n-1$ variables and once
more to use the final rescaled arguments.

\section{Application to the Reduction to Basis Integrals}
\label{sec:application}

Let us now assess the efficiency of the algorithm presented in
section~\ref{sec:multi_rec} in practical applications. For brevity, we
will refer to the new method as ``scaling'' reconstruction. We compare
it to an algorithm proposed by Cuyt and Lee~\cite{CuytL11}. This
algorithm is described in detail
in~\cite{Peraro:2016wsq,Klappert:2019emp}. We briefly recall the main
steps. First, the variables are rescaled with a common factor $t$ and
shifted, leading to $(x_1,\dots,x_n)$ = $(ty_1+s_1,\dots,ty_n+s_n)$
with $y_n = 1$. One then performs a univariate rational reconstruction
in $t$. Starting from the highest powers, the coefficients of $t^i$
are reconstructed as polynomials in $y_1,\dots,y_n$ and transformed
back to the original variables. For the comparison we use
state-of-the-art implementations in the public codes
\texttt{FireFly}~\cite{Klappert:2019emp,Klappert:2020aqs,Firefly} and
\texttt{FiniteFlow}~\cite{Peraro:2019svx,FiniteFlow}. The comparison
code and the example rational functions in computer-readable form are
available from \url{https://github.com/a-maier/scaling-rec}.

For the scaling algorithm, the reconstruction is first performed over
a number of prime fields $\mathbb{Z}_P$, using arithmetic algorithms
taken from NTL~\cite{NTL}. We start with $P=\num{1152921504606846883}$
and move to the next smaller prime numbers as needed. The resulting
coefficients are lifted to a higher characteristic with the Chinese
remainder theorem, specifically Bézout's identity. The actual rational
coefficients are then reconstructed from the finite-field integers via
an algorithm by Wang~\cite{Wang}. Again, details are given
in~\cite{Peraro:2016wsq,Klappert:2019emp}.

In principle, the reconstruction can be simplified tremendously
exploiting the structure of the
result~\cite{Abreu:2018zmy,DeLaurentis:2022otd,Chawdhry:2023yyx}. Both
\texttt{FireFly} and \texttt{FiniteFlow} can factorise the numerator
and denominator to a certain degree. \texttt{FiniteFlow} determines
the minimal degree in each variable to automatically factor out common
monomials. \texttt{FireFly} optionally performs univariate
factorisation to identify any factors depending on a single
variable. Since we are mainly interested in assessing the underlying
reconstruction algorithms, we disable \texttt{FireFly}'s factorisation
in the following comparisons. As there is no option to switch off
factorisation in \texttt{FiniteFlow}, we additionally compare the
reconstruction after removing all monomial factors from the function
to be reconstructed.

\subsection{Massive Four-Loop Propagator}
\label{sec:4l}

Our first benchmark point is a coefficient in the differential
equation~\cite{Kotikov:1990kg,Remiddi:1997ny} for a four-loop massive
propagator. Setting the mass $m=1$, the propagator is a function of
$z=p^2$, where $p$ is the external four-momentum. One obtains
\begin{equation}
  z \frac{d}{dz}\ \centeredgraphics{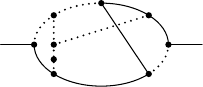} =  q(z, d) \ \centeredgraphics{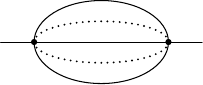} + \dots,
\end{equation}
with the ellipsis indicating a linear combination of further basis
integrals with less complex coefficients. $q(z, d)$ is a rational
function in $z$ and the space-time dimension $d$, where the numerator
degrees in $z$ and $d$ are $P_z = P_d = 81$ and the denominator
degrees are $Q_z = 80$ and $Q_d = 78$. The numbers of probes required
for reconstructing the function over the first characteristic are
shown in table~\ref{tab:probes_prop} for the scaling algorithm and
\texttt{FireFly} together with a hypothetical optimal algorithm that
can determine one unknown coefficient per probe.

For a full rational function reconstruction probes are needed in
several additional prime fields. Since different implementations vary
vastly in the amount of reused information we refrain from a
quantitative comparison.

\begin{table}[htb]
  \centering
  \begin{tabular}{ll}
    \toprule
    Method & Number of Probes\\
    \midrule
    This work & $\phantom{\gtrsim{}}\num{13594}$\\
    \texttt{FireFly} & $\phantom{\gtrsim{}}\num{16373}$\\
    Optimal & $\phantom{\gtrsim{}}\num{12721}$\\
    \bottomrule
  \end{tabular}
  \caption{%
    Number of probes required to reconstruct a specific coefficient in
    the differential equation for a four-loop propagator over the
    first characteristic.
  }
  \label{tab:probes_prop}
\end{table}

Comparing the ``Optimal'' entry of table~\ref{tab:probes_prop} to the
total number of \num{13123} monomials in the ansatz given by
equation~\eqref{f_multivar} we observe that $q(z, d)$ is dense in the
sense that about $97\%$ of the coefficients in the ansatz are
non-zero. We see that the scaling algorithm performs close to optimal,
with an overhead of about $7\%$ additional probes. In comparison,
\texttt{FireFly} requires approximately $29\%$ more probes.

The denominator of the reconstructed function contains an overall
factor of $d^1z^3$. The improvement gained by identifying and removing
this factor is illustrated in table~\ref{tab:probes_prop_mod}, where
we now also include \texttt{FiniteFlow}. The \texttt{FiniteFlow} entry
does not include a few hundred probes used to determine the overall
degree of the rational function and the degrees with respect to the
individual variables from Thiele interpolation, c.f.\
section~\ref{sec:uni_rec}~\cite{FiniteFlowMiss}. The other entries
count the total number of function evaluations.

\begin{table}[htb]
  \centering
  \begin{tabular}{ll}
    \toprule
    Method & Number of Probes\\
    \midrule
    This work & $\phantom{\gtrsim{}}\num{13594}$\\
    \texttt{FireFly} & $\phantom{\gtrsim{}}\num{16020}$\\
    \texttt{FiniteFlow} & $\gtrsim{}\num{18205}$\\
    Optimal & $\phantom{\gtrsim{}}\num{12721}$\\
    \bottomrule
  \end{tabular}
  \caption{%
    Number of probes required to reconstruct a specific coefficient in
    the differential equation for a four-loop propagator over the
    first characteristic after removing an overall monomial
    factor. The \texttt{FiniteFlow} entry does not include a few
    hundred probes used for degree determinations.
  }
  \label{tab:probes_prop_mod}
\end{table}

For the algorithm presented in section~\ref{sec:multi_rec}, the
scaling powers in equation~\eqref{eq:scaling_rec} are completely
determined by the numerator in the present example. Thus, removing
factors from the denominator does not affect the number of probes
needed. However, we do observe a slight reduction in the number of
required evaluations with $\texttt{FireFly}$, reducing the overhead to
$26\%$ compared to the optimum. The number of evaluations needed with
$\texttt{FiniteFlow}$ exceeds the number of non-vanishing coefficients
to be determined by about $43\%$.

\subsection{Diphoton Plus Jet Production at Two Loops}
\label{sec:diphoton_jet}

Next, let us consider the two-loop amplitude for diphoton plus jet
production, taken from~\cite{Agarwal:2021vdh}. Specifically, we choose
the parity-even contribution with a left-handed quark and a gluon in
the initial state, a negative gluon helicity, opposite-sign photon
helicities, and no closed fermion loops. Denoting the quark helicity
by $\lambda_q$, the number of active flavours by $n_f$, the number of
colours by $N_C$, and the parity transformation operator by $\hat{P}$,
the reduction has the structure
\begin{equation}
  \frac{1 + \hat{P}}{2}\ \left[\centeredgraphics{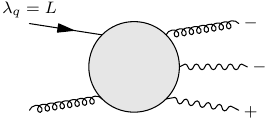}\right]_{n_f = 0} = c_{-2} N_C^{-2} + c_{0}N_C^0 + c_{2} N_C^2,
\end{equation}
where $c_{-2}, c_{0}, c_{2}$ are linear combinations of pentagon
functions~\cite{Gehrmann:2018yef} with rational coefficients. From
$c_{0}$ we select the largest of these coefficients by Mathematica's
\texttt{ByteCount}. We write this coefficient as a rational function
in $x_{23},x_{34},x_{45},x_{51}$, where
$x_{ij} = \frac{s_{ij}}{s_{12}}$, and
\begin{equation*}
  s_{12} = (p_1+p_2)^2,  s_{23} = (p_2-p_3)^2,  s_{34} = (p_3+p_4)^2,  s_{45} = (p_4+p_5)^2,  s_{51} = (p_1-p_5)^2
\end{equation*}
are Mandelstam invariants. After determining the numerator and
denominator degrees our ansatz according to
equation~\eqref{f_multivar} contains \num{136934} unknown
coefficients. However, the actual rational function is much sparser
than in the example in section~\ref{sec:4l} and only approximately
$22\%$ of these coefficients are non-zero. In this example, the full
coefficient can be reconstructed using a single prime field. We
collect the number of required probes in
table~\ref{tab:probes_diphoton_jet}.

\begin{table}[htb]
  \centering
  \begin{tabular}{ll}
    \toprule
    Method & Number of Probes\\
    \midrule
    This work & $\phantom{\gtrsim{}}\num{169132}$ \\
    \texttt{FireFly} & $\phantom{\gtrsim{}}\num{163094}$ \\
    Optimal & $\phantom{\gtrsim{}}\num{30490}$\\
    \bottomrule
  \end{tabular}
  \caption{Number of probes required to reconstruct the coefficient in
    the reduction of the two-loop diphoton plus jet amplitude.}
  \label{tab:probes_diphoton_jet}
\end{table}

The scaling algorithm introduced in section~\ref{sec:multi_rec}
performs slightly worse than \texttt{FireFly}'s reconstruction. Both
algorithms are far from optimal for this scenario, requiring more than
five probes for each unknown coefficient.

The number of required reconstruction probes after removing an overall
monomial factor $x_{23}^2x_{34}^2x_{45}^2x_{51}^2$ is shown in table
\ref{tab:probes_diphoton_jet_mod}. As in section~\ref{sec:4l}, we see
no improvement for the implementation of the algorithm presented in
this work. In contrast, \texttt{FireFly} needs approximately $20\%$
fewer function evaluations than before. Most strikingly,
\texttt{FiniteFlow} is much closer to optimal than both
\texttt{FireFly} and the scaling reconstruction implementation,
especially when using the \texttt{FFPolyVandermonde} alternative
polynomial reconstruction method. Even when enabling its
identification of univariate factors, \texttt{FireFly} still requires
\num{87485} probes, substantially more than \texttt{FiniteFlow}. This
difference between \texttt{FiniteFlow} and \texttt{FireFly} is
unexpected and deserves closer inspection. However, since the focus of
the present work is on the scaling algorithm and dense reconstruction,
we leave further investigation to future work.

\begin{table}[htb]
  \centering
  \begin{center}
    \begin{tabular}{ll}
      \toprule
      Method & Number of Probes\\
      \midrule
      This work & $\phantom{\gtrsim{}}\num{169132}$ \\
      \texttt{FireFly} & $\phantom{\gtrsim{}}\num{129894}$ \\
      \texttt{FiniteFlow} & $\gtrsim\num{49216}$\\
      \texttt{FiniteFlow} with \texttt{FFPolyVandermonde} & $\gtrsim\num{47381}$\\
      Optimal & $\phantom{\gtrsim{}}\num{30490}$\\
      \bottomrule
    \end{tabular}
  \end{center}
  \caption{%
    Number of required probes after removing the overall monomial
    prefactor. The \texttt{FiniteFlow} entries do not include a few
    hundred probes used for degree determinations.}
  \label{tab:probes_diphoton_jet_mod}
\end{table}

\section{Conclusion}
\label{sec:conclusion}

I have presented an algorithm for the reconstruction of dense
multivariate rational functions. Multiple variables are mapped onto a
single variable, using scaling powers and shifts chosen such that the
mapping can be inverted. In this way, the problem is reduced to
well-known univariate rational reconstruction.

The algorithm is tested on two examples taken from complex reductions
to basis integrals, a massive four-loop propagator and a two-loop
five-point amplitude. For the dense rational function encountered in
the four-loop problem, the required number of probes exceeds the
number of unknown coefficients by only about $7\%$. This compares
favourably with the current state-of-the-art programs
\texttt{FireFly}~\cite{Klappert:2019emp,Klappert:2020aqs} and
\texttt{FiniteFlow}~\cite{Peraro:2019svx}.

In the sparse two-loop example, the number of probes needed is about
$4\%$ above the \texttt{FireFly} result when disabling
factorisation. However, a comparison to \texttt{FiniteFlow} reveals
that in this case there is substantial room for improvements for both
\texttt{FireFly} and the presented algorithm. A further promising
avenue for future research would be to combine the univariate mapping
with sparse rational reconstruction in a single variable, see
e.g.~\cite{Huang:2017}.

\subsection*{Acknowledgements}
\label{ack}

I thank P.~Marquard for enlightening discussions and helpful comments
on the manuscript. I further thank F.~Lange and T.~Peraro for
communication on \texttt{FireFly} and \texttt{FiniteFlow} and for
contributing code to the comparison in section~\ref{sec:application}.

\bibliographystyle{JHEP}
\bibliography{biblio.bib}

\providecommand{\href}[2]{#2}\begingroup\raggedright\begin{thebibliography}{10}

\bibitem{Chetyrkin:1981qh}
K.~G. Chetyrkin and F.~V. Tkachov, \emph{{Integration by parts: The algorithm
  to calculate $\beta$-functions in 4 loops}},
  \href{https://doi.org/10.1016/0550-3213(81)90199-1}{\emph{Nucl. Phys. B}
  {\bfseries 192} (1981) 159}.

\bibitem{Tkachov:1981wb}
F.~V. Tkachov, \emph{{A theorem on analytical calculability of 4-loop
  renormalization group functions}},
  \href{https://doi.org/10.1016/0370-2693(81)90288-4}{\emph{Phys. Lett. B}
  {\bfseries 100} (1981) 65}.

\bibitem{Laporta:2001dd}
S.~Laporta, \emph{{High precision calculation of multiloop Feynman integrals by
  difference equations}}, \href{https://doi.org/10.1016/S0217-751X(00)00215-7,
  10.1142/S0217751X00002157}{\emph{Int. J. Mod. Phys.} {\bfseries A15} (2000)
  5087} [\href{https://arxiv.org/abs/hep-ph/0102033}{{\ttfamily
  hep-ph/0102033}}].

\bibitem{Kauers:2008zz}
M.~Kauers, \emph{{Fast solvers for dense linear systems}},
  \href{https://doi.org/10.1016/j.nuclphysbps.2008.09.111}{\emph{Nucl. Phys. B
  Proc. Suppl.} {\bfseries 183} (2008) 245}.

\bibitem{Kant:2013vta}
P.~Kant, \emph{{Finding Linear Dependencies in Integration-By-Parts Equations:
  A Monte Carlo Approach}},
  \href{https://doi.org/10.1016/j.cpc.2014.01.017}{\emph{Comput. Phys. Commun.}
  {\bfseries 185} (2014) 1473}
  [\href{https://arxiv.org/abs/1309.7287}{{\ttfamily 1309.7287}}].

\bibitem{vonManteuffel:2014ixa}
A.~von Manteuffel and R.~M. Schabinger, \emph{{A novel approach to integration
  by parts reduction}},
  \href{https://doi.org/10.1016/j.physletb.2015.03.029}{\emph{Phys. Lett. B}
  {\bfseries 744} (2015) 101}
  [\href{https://arxiv.org/abs/1406.4513}{{\ttfamily 1406.4513}}].

\bibitem{Peraro:2016wsq}
T.~Peraro, \emph{{Scattering amplitudes over finite fields and multivariate
  functional reconstruction}},
  \href{https://doi.org/10.1007/JHEP12(2016)030}{\emph{JHEP} {\bfseries 12}
  (2016) 030} [\href{https://arxiv.org/abs/1608.01902}{{\ttfamily
  1608.01902}}].

\bibitem{Gorishnii:1989gt}
S.~G. Gorishnii, S.~A. Larin, L.~R. Surguladze and F.~V. Tkachov,
  \emph{{Mincer: Program for Multiloop Calculations in Quantum Field Theory for
  the Schoonschip System}},
  \href{https://doi.org/10.1016/0010-4655(89)90134-3}{\emph{Comput. Phys.
  Commun.} {\bfseries 55} (1989) 381}.

\bibitem{Steinhauser:2000ry}
M.~Steinhauser, \emph{{MATAD: A Program package for the computation of MAssive
  TADpoles}},
  \href{https://doi.org/10.1016/S0010-4655(00)00204-6}{\emph{Comput. Phys.
  Commun.} {\bfseries 134} (2001) 335}
  [\href{https://arxiv.org/abs/hep-ph/0009029}{{\ttfamily hep-ph/0009029}}].

\bibitem{Gluza:2010ws}
J.~Gluza, K.~Kajda and D.~A. Kosower, \emph{{Towards a Basis for Planar
  Two-Loop Integrals}},
  \href{https://doi.org/10.1103/PhysRevD.83.045012}{\emph{Phys. Rev. D}
  {\bfseries 83} (2011) 045012}
  [\href{https://arxiv.org/abs/1009.0472}{{\ttfamily 1009.0472}}].

\bibitem{Schabinger:2011dz}
R.~M. Schabinger, \emph{{A New Algorithm For The Generation Of
  Unitarity-Compatible Integration By Parts Relations}},
  \href{https://doi.org/10.1007/JHEP01(2012)077}{\emph{JHEP} {\bfseries 01}
  (2012) 077} [\href{https://arxiv.org/abs/1111.4220}{{\ttfamily 1111.4220}}].

\bibitem{Lee:2012cn}
R.~N. Lee, \emph{{Presenting LiteRed: a tool for the Loop InTEgrals
  REDuction}},  \href{https://arxiv.org/abs/1212.2685}{{\ttfamily 1212.2685}}.

\bibitem{Lee:2013mka}
R.~N. Lee, \emph{{LiteRed 1.4: a powerful tool for reduction of multiloop
  integrals}}, \href{https://doi.org/10.1088/1742-6596/523/1/012059}{\emph{J.
  Phys. Conf. Ser.} {\bfseries 523} (2014) 012059}
  [\href{https://arxiv.org/abs/1310.1145}{{\ttfamily 1310.1145}}].

\bibitem{Ita:2015tya}
H.~Ita, \emph{{Two-loop Integrand Decomposition into Master Integrals and
  Surface Terms}},
  \href{https://doi.org/10.1103/PhysRevD.94.116015}{\emph{Phys. Rev. D}
  {\bfseries 94} (2016) 116015}
  [\href{https://arxiv.org/abs/1510.05626}{{\ttfamily 1510.05626}}].

\bibitem{Larsen:2015ped}
K.~J. Larsen and Y.~Zhang, \emph{{Integration-by-parts reductions from
  unitarity cuts and algebraic geometry}},
  \href{https://doi.org/10.1103/PhysRevD.93.041701}{\emph{Phys. Rev. D}
  {\bfseries 93} (2016) 041701}
  [\href{https://arxiv.org/abs/1511.01071}{{\ttfamily 1511.01071}}].

\bibitem{Ruijl:2017cxj}
B.~Ruijl, T.~Ueda and J.~A.~M. Vermaseren, \emph{{Forcer, a FORM program for
  the parametric reduction of four-loop massless propagator diagrams}},
  \href{https://doi.org/10.1016/j.cpc.2020.107198}{\emph{Comput. Phys. Commun.}
  {\bfseries 253} (2020) 107198}
  [\href{https://arxiv.org/abs/1704.06650}{{\ttfamily 1704.06650}}].

\bibitem{Pikelner:2017tgv}
A.~Pikelner, \emph{{FMFT: Fully Massive Four-loop Tadpoles}},
  \href{https://doi.org/10.1016/j.cpc.2017.11.017}{\emph{Comput. Phys. Commun.}
  {\bfseries 224} (2018) 282}
  [\href{https://arxiv.org/abs/1707.01710}{{\ttfamily 1707.01710}}].

\bibitem{Wu:2023upw}
Z.~Wu, J.~Boehm, R.~Ma, H.~Xu and Y.~Zhang, \emph{{NeatIBP 1.0, a package
  generating small-size integration-by-parts relations for Feynman integrals}},
  \href{https://doi.org/10.1016/j.cpc.2023.108999}{\emph{Comput. Phys. Commun.}
  {\bfseries 295} (2024) 108999}
  [\href{https://arxiv.org/abs/2305.08783}{{\ttfamily 2305.08783}}].

\bibitem{Guan:2024byi}
X.~Guan, X.~Liu, Y.-Q. Ma and W.-H. Wu, \emph{{Blade: A package for
  block-triangular form improved Feynman integrals decomposition}},
  \href{https://doi.org/10.1016/j.cpc.2025.109538}{\emph{Comput. Phys. Commun.}
  {\bfseries 310} (2025) 109538}
  [\href{https://arxiv.org/abs/2405.14621}{{\ttfamily 2405.14621}}].

\bibitem{Magerya:2022hvj}
V.~Magerya, \emph{{Rational Tracer: a Tool for Faster Rational Function
  Reconstruction}},  \href{https://arxiv.org/abs/2211.03572}{{\ttfamily
  2211.03572}}.

\bibitem{Smirnov:2020quc}
A.~V. Smirnov and V.~A. Smirnov, \emph{{How to choose master integrals}},
  \href{https://doi.org/10.1016/j.nuclphysb.2020.115213}{\emph{Nucl. Phys. B}
  {\bfseries 960} (2020) 115213}
  [\href{https://arxiv.org/abs/2002.08042}{{\ttfamily 2002.08042}}].

\bibitem{Usovitsch:2020jrk}
J.~Usovitsch, \emph{{Factorization of denominators in integration-by-parts
  reductions}},  \href{https://arxiv.org/abs/2002.08173}{{\ttfamily
  2002.08173}}.

\bibitem{Abreu:2018zmy}
S.~Abreu, J.~Dormans, F.~Febres~Cordero, H.~Ita and B.~Page, \emph{{Analytic
  Form of Planar Two-Loop Five-Gluon Scattering Amplitudes in QCD}},
  \href{https://doi.org/10.1103/PhysRevLett.122.082002}{\emph{Phys. Rev. Lett.}
  {\bfseries 122} (2019) 082002}
  [\href{https://arxiv.org/abs/1812.04586}{{\ttfamily 1812.04586}}].

\bibitem{DeLaurentis:2022otd}
G.~De~Laurentis and B.~Page, \emph{{Ans\"atze for scattering amplitudes from
  p-adic numbers and algebraic geometry}},
  \href{https://doi.org/10.1007/JHEP12(2022)140}{\emph{JHEP} {\bfseries 12}
  (2022) 140} [\href{https://arxiv.org/abs/2203.04269}{{\ttfamily
  2203.04269}}].

\bibitem{Chawdhry:2023yyx}
H.~A. Chawdhry, \emph{{p-adic reconstruction of rational functions in multiloop
  amplitudes}}, \href{https://doi.org/10.1103/PhysRevD.110.056028}{\emph{Phys.
  Rev. D} {\bfseries 110} (2024) 056028}
  [\href{https://arxiv.org/abs/2312.03672}{{\ttfamily 2312.03672}}].

\bibitem{Liu:2023cgs}
X.~Liu, \emph{{Reconstruction of rational functions made simple}},
  \href{https://doi.org/10.1016/j.physletb.2024.138491}{\emph{Phys. Lett. B}
  {\bfseries 850} (2024) 138491}
  [\href{https://arxiv.org/abs/2306.12262}{{\ttfamily 2306.12262}}].

\bibitem{Klappert:2019emp}
J.~Klappert and F.~Lange, \emph{{Reconstructing rational functions with
  FireFly}}, \href{https://doi.org/10.1016/j.cpc.2019.106951}{\emph{Comput.
  Phys. Commun.} {\bfseries 247} (2020) 106951}
  [\href{https://arxiv.org/abs/1904.00009}{{\ttfamily 1904.00009}}].

\bibitem{Peraro:2019svx}
T.~Peraro, \emph{{$\text{FiniteFlow}$: multivariate functional reconstruction
  using finite fields and dataflow graphs}},
  \href{https://doi.org/10.1007/JHEP07(2019)031}{\emph{JHEP} {\bfseries 07}
  (2019) 031} [\href{https://arxiv.org/abs/1905.08019}{{\ttfamily
  1905.08019}}].

\bibitem{Klappert:2020aqs}
J.~Klappert, S.~Y. Klein and F.~Lange, \emph{{Interpolation of dense and sparse
  rational functions and other improvements in FireFly}},
  \href{https://doi.org/10.1016/j.cpc.2021.107968}{\emph{Comput. Phys. Commun.}
  {\bfseries 264} (2021) 107968}
  [\href{https://arxiv.org/abs/2004.01463}{{\ttfamily 2004.01463}}].

\bibitem{Belitsky:2023qho}
A.~V. Belitsky, A.~V. Smirnov and R.~V. Yakovlev, \emph{{Balancing act:
  Multivariate rational reconstruction for IBP}},
  \href{https://doi.org/10.1016/j.nuclphysb.2023.116253}{\emph{Nucl. Phys. B}
  {\bfseries 993} (2023) 116253}
  [\href{https://arxiv.org/abs/2303.02511}{{\ttfamily 2303.02511}}].

\bibitem{thiele}
T.~N. Thiele, \emph{Interpolationsrechnung}. B. G. Teubner, 1909.

\bibitem{Agarwal:2021vdh}
B.~Agarwal, F.~Buccioni, A.~von Manteuffel and L.~Tancredi, \emph{{Two-Loop
  Helicity Amplitudes for Diphoton Plus Jet Production in Full Color}},
  \href{https://doi.org/10.1103/PhysRevLett.127.262001}{\emph{Phys. Rev. Lett.}
  {\bfseries 127} (2021) 262001}
  [\href{https://arxiv.org/abs/2105.04585}{{\ttfamily 2105.04585}}].

\bibitem{CuytL11}
A.~A.~M. Cuyt and W.~Lee, \emph{Sparse interpolation of multivariate rational
  functions}, \href{https://doi.org/10.1016/J.TCS.2010.11.050}{\emph{Theor.
  Comput. Sci.} {\bfseries 412} (2011) 1445}.

\bibitem{Firefly}
J.~Klappert, S.~Y. Klein and F.~Lange, {FireFly version 2.0.3,
  \url{https://gitlab.com/firefly-library/firefly}}.

\bibitem{FiniteFlow}
T.~Peraro, {$\text{FiniteFlow}$ revision
  7754b861168ae5aea4c4d555a37e2e88d18b7725,
  \url{https://github.com/peraro/finiteflow}}.

\bibitem{NTL}
V.~Shoup, \emph{NTL: A Library for doing Number Theory}. \url{https://libntl.org/}

\bibitem{Wang}
P.~S. Wang, \emph{A p-adic algorithm for univariate partial fractions},  in
  \emph{Proceedings of the Fourth ACM Symposium on Symbolic and Algebraic
  Computation}, SYMSAC '81, (New York, NY, USA), p.~212–217, Association for
  Computing Machinery, 1981, \href{https://doi.org/10.1145/800206.806398}{DOI}.

\bibitem{Kotikov:1990kg}
A.~V. Kotikov, \emph{{Differential equations method: New technique for massive
  Feynman diagrams calculation}},
  \href{https://doi.org/10.1016/0370-2693(91)90413-K}{\emph{Phys. Lett. B}
  {\bfseries 254} (1991) 158}.

\bibitem{Remiddi:1997ny}
E.~Remiddi, \emph{{Differential equations for Feynman graph amplitudes}},
  \href{https://doi.org/10.1007/BF03185566}{\emph{Nuovo Cim. A} {\bfseries 110}
  (1997) 1435} [\href{https://arxiv.org/abs/hep-th/9711188}{{\ttfamily
  hep-th/9711188}}].

\bibitem{FiniteFlowMiss}
T.~Peraro, Private communication.

\bibitem{Gehrmann:2018yef}
T.~Gehrmann, J.~M. Henn and N.~A. Lo~Presti, \emph{{Pentagon functions for
  massless planar scattering amplitudes}},
  \href{https://doi.org/10.1007/JHEP10(2018)103}{\emph{JHEP} {\bfseries 10}
  (2018) 103} [\href{https://arxiv.org/abs/1807.09812}{{\ttfamily
  1807.09812}}].

\bibitem{Huang:2017}
Q.~Huang and X.~Gao, \emph{Sparse rational function interpolation with finitely
  many values for the coefficients}, {\emph{CoRR} {\bfseries abs/1706.00914}
  (2017) } [\href{https://arxiv.org/abs/1706.00914}{{\ttfamily 1706.00914}}].

\end{thebibliography}\endgroup

\end{document}